\documentstyle[pra,preprint,aps]{revtex}
\draft

\newcommand{\ba}{\begin{eqnarray}}
\newcommand{\ea}{\end{eqnarray}}
\begin{document}
\pagestyle{plain}

\title{Comment on ``Boson-realization model for the vibrational 
spectra of tetrahedral molecules''}

\author{R. Lemus$^{1)}$, F. P\'erez-Bernal${^2)}$, A. Frank$^{1,3)}$, 
R. Bijker$^{1)}$ and J.M. Arias$^{2)}$}
\address{
$^{1)}$ Instituto de Ciencias Nucleares, 
        Universidad Nacional Aut\'onoma de M\'exico, 
        A.P. 70-543, 04510 M\'exico D.F., M\'exico\\
$^{2)}$ Departamento de F\'{\i}sica At\'omica, Molecular y Nuclear,
        Facultad de F\'{\i}sica, Universidad de Sevilla,
        Apdo. 1065, 41080 Sevilla, Espa\~na\\
$^{3)}$ Instituto de F\'{\i}sica, Laboratorio de Cuernavaca,
        A.P. 139-B, Cuernavaca, Morelos, M\'exico}
\maketitle

\begin{abstract}
An algebraic model in terms of a local harmonic boson realization 
was recently proposed to study molecular vibrational spectra 
[Zhong-Qi Ma et al., Phys. Rev. A {\bf 53}, 2173 (1996)]. 
Because of the local nature of the bosons the model has to deal 
with spurious degrees of freedom. An approach to eliminate the 
latter from both the Hamiltonian and the basis was suggested. 
We show that this procedure does not remove all spurious components 
from the Hamiltonian and leads to a restricted set of interactions. 
We then propose a scheme in which the physical Hamiltonian can be  
systematically constructed up to any order without the need of 
imposing conditions on its matrix elements. In addition, we show 
that this scheme corresponds to the harmonic limit
of a symmetry adapted algebraic approach based on $U(2)$ algebras.
\end{abstract}

\begin{center}
\end{center}

\clearpage
\section{Introduction}

In a recent paper \cite{Ma}, Zhong-Qi Ma et al. proposed a model 
to describe the vibrational spectra of tetrahedral molecules and 
applied it to the methane molecule. The authors assign to each of 
the ten interatomic bonds a set of bosonic operators 
$\{ a^{\dagger}_i, a_i \}$, $i=1,2,\ldots,10$~, 
which satisfy the standard boson commutation relations
\ba
\left[ a_i,a^{\dagger}_j \right] \;=\; \delta_{i,j} ~, 
\hspace{1cm} \left[ a_i,a_j \right] \;=\; 
\left[ a^{\dagger}_i,a^{\dagger}_j \right] \;=\; 0 ~.
\ea
The first four operators ($i=1,\ldots,4$) are chosen to correspond 
to the C-H bonds (stretching vibrations), while the last six 
($i=5,\ldots,10$) represent the H-H bonds (bending vibrations). 
The Hamiltonian is then expanded in terms of these operators. For 
methane they proposed the following ${\cal T}_d$ invariant Hamiltonian 
\ba
\hat H &=& \hat H_1 + \hat H_{2s} + \hat H_{2b} ~, 
\label{ham} 
\ea
where $\hat H_1$ is a one-body contribution 
\ba
\hat H_1 &=& \epsilon_s \sum_{i=1}^{4} a^{\dagger}_i a_i 
+ \epsilon_b \sum_{i=5}^{10} a^{\dagger}_i a_i  
+ \lambda_1 \sum_{i \neq j=1}^{4}  a^{\dagger}_i a_j 
+ \lambda_2 \sum_{i \neq j=5}^{10} a^{\dagger}_i a_j
\nonumber\\
&& + \lambda_3 \left[ (a^{\dagger}_5 a_8 + a^{\dagger}_6 a_9 
+ a^{\dagger}_7 a_{10}) + \mbox{h.c.} \right] 
\nonumber\\
&& + \lambda_4 \left[ a^{\dagger}_1 (a_5 + a_6 + a_7 - a_8 - a_9 -a_{10}) 
\right.
\nonumber\\
&& \hspace{1cm} 
+ a^{\dagger}_2 (  a_5 - a_6 - a_7 - a_8 + a_9 + a_{10}) 
\nonumber\\ 
&& \hspace{1cm}
+ a^{\dagger}_3 (- a_5 + a_6 - a_7 + a_8 - a_9 + a_{10}) 
\nonumber\\ 
&& \hspace{1cm} \left.
+ a^{\dagger}_4 (- a_5 - a_6 + a_7 + a_8 + a_9 - a_{10}) 
+ \mbox{h.c.} \right] 
\nonumber\\  
&& + \lambda_5 \left[ (\sum_{i=1}^{4} a^{\dagger}_i) 
(\sum_{j=5}^{10} a_j) + \mbox{h.c.} \right] ~,
\label{ham1}
\ea
and $\hat H_{2x}$ corresponds to two-body anharmonic contributions 
to the stretching ($x=s$) and bending ($x=b$) vibrations 
\ba
\hat H_{2s} &=& -X_s \sum_{i=1}^{4}  \hat N_i (\hat N_i - 1) 
\;=\; -X_s \sum_{i=1}^{4} a^{\dagger}_i a^{\dagger}_i a_i a_i ~, 
\nonumber\\
\hat H_{2b} &=& -X_b \sum_{i=5}^{10} \hat N_i (\hat N_i - 1) 
\;=\; -X_b \sum_{i=5}^{10} a^{\dagger}_i a^{\dagger}_i a_i a_i ~. 
\label{ham2}
\ea
In Eq.~(\ref{ham1}) we introduced $\epsilon_s=\omega_s-2X_s$ 
and $\epsilon_b=\omega_b-2X_b$. The number operator $\hat N_i$ is 
defined as $\hat N_i=a^{\dagger}_i a_i$. The Hamiltonian conserves 
the total number of quanta $N=\sum_{i=1}^{10} N_i$. 
Since the methane molecule has nine vibrational degrees of freedom, 
while the operators $\{ a^{\dagger}_i, a_i \}$ give 
rise to ten degrees of freedom, spurious interactions may occur 
in $\hat H$. Their identification, however, is not obvious when the
Hamiltonian is written in the form of Eq.~(\ref{ham})--(\ref{ham2}). 
In \cite{Ma} the authors proceed by identifying the spurious states 
as those containing the fully symmetric, one-quantum bending state 
$\psi(A_1;100000)$ as a factor. With a combination of 
projection and coupling techniques and an orthogonalization 
procedure, the physical space up to three quanta is generated. 
Next it is shown that the physical Hamiltonian does not couple 
the spurious states with the physical states as long as the conditions 
\ba
\lambda_5 \;=\; 0 ~, \hspace{1cm} X_b \;=\; 0 ~, 
\label{cond1} 
\ea
are imposed on the Hamiltonian of Eq.~(\ref{ham}). 
Although these conditions indeed lead to the correct results, they 
are too restrictive and, furthermore, do not remove all spurious 
interactions. 

\section{Spurious interactions}

In this section, we show that the condition of Eq.~(\ref{cond1}) is 
not sufficient to remove all spurious interactions from the Hamiltonian. 
We define a new set of harmonic bosons with well defined tensorial 
properties under the tetrahedral ${\cal T}_d$ group by projecting the 
local boson operators. For the stretching modes we have 
\ba
s^{\dagger}_{\Gamma,\gamma} &=& 
\sum_{i=1}^{4} \, \alpha^{\Gamma,\gamma}_i \, a^{\dagger}_i ~,
\label{stretch}
\ea
with $\Gamma=A_1$, $F_2$ and for the bending modes
\ba
b^{\dagger}_{\Gamma,\gamma} &=& 
\sum_{i=5}^{10} \, \beta^{\Gamma,\gamma}_i \, a^{\dagger}_i ~, 
\label{bend}
\ea
with $\Gamma=A_1$, $E$, $F_2$, and similar expressions for the 
annihilation operators. The coefficients $\alpha^{\Gamma,\gamma}_i$ 
and $\beta^{\Gamma,\gamma}_i$ are the same as those appearing in the 
projected wave functions of Eq.~(3.1-2) in \cite{Ma}.
In the framework of this alternative scheme, both the Hamiltonian 
and the basis states are constructed by repeated coupling of these 
normal bosons. In this way, we obtain automatically an orthonormal 
symmetry adapted basis with a well defined number of quanta in 
each one of the fundamental vibrational modes. For example, 
with the usual notation for the normal modes of methane 
($\nu_1,\nu_2,\nu_3,\nu_4$), we find for two quanta the following 
basis states with symmetry $A_1$ 
\ba
| 2,0,0,0 \rangle &=& \frac{1}{\sqrt{2}} 
(s^{\dagger}_{A_1} s^{\dagger}_{A_1})^{A_1} | 0 \rangle ~,
\nonumber\\
| 0,2,0,0 \rangle &=& \frac{1}{\sqrt{2}} 
(b^{\dagger}_{E} b^{\dagger}_{E})^{A_1} | 0 \rangle ~,
\nonumber\\
| 0,0,2,0 \rangle &=& \frac{1}{\sqrt{2}} 
(s^{\dagger}_{F_2} s^{\dagger}_{F_2})^{A_1} | 0 \rangle ~,
\nonumber\\
| 0,0,0,2 \rangle &=& \frac{1}{\sqrt{2}} 
(b^{\dagger}_{F_2} b^{\dagger}_{F_2})^{A_1} | 0 \rangle ~.
\nonumber\\
| 0,0,1,1 \rangle &=&  
(s^{\dagger}_{F_2} b^{\dagger}_{F_2})^{A_1} | 0 \rangle ~,
\label{basis}
\ea
Here $(\ldots)^{\Gamma}$ indicates coupling of two quanta to 
symmetry $\Gamma$. 
These basis states are orthogonal, since they correspond to eigenfunctions 
associated with different eigenvalues of the number operators 
\ba
\hat n_{\Gamma_s} \;=\; 
\sum_{\gamma} s^{\dagger}_{\Gamma,\gamma} s_{\Gamma,\gamma} ~, 
\hspace{1cm} \hat n_{\Gamma_b} \;=\; 
\sum_{\gamma} b^{\dagger}_{\Gamma,\gamma} b_{\Gamma,\gamma} ~. 
\label{number}
\ea
This symmetry adapted procedure provides a method to remove the spurious 
interactions from the Hamiltonian as well as the spurious states from 
the basis. It consists in deleting the tensors which are associated 
with the spurious bending mode, $b^{\dagger}_{A_1}$ and $b_{A_1}$, 
in the construction of physical operators and basis states. This allows 
an exact elimination of the spurious states without the need to 
impose additional conditions on matrix elements. 

The Hamiltonian of Eq.~(\ref{ham}) does include spurious interactions. 
This can be seen by rewriting $\hat H$ in terms of the normal bosons 
of Eqs.~(\ref{stretch}) and~(\ref{bend}). For the one-body Hamiltonian 
of Eq.~(\ref{ham1}) we find 
\ba
\hat H_1 &=& (\epsilon_s+3\lambda_1) \, \hat n_{A_{1s}} 
+ (\epsilon_s-\lambda_1) \, \hat n_{F_{2s}} 
\nonumber\\
&&+ (\epsilon_b+5\lambda_2+\lambda_3) \, \hat n_{A_{1b}} 
  + (\epsilon_b- \lambda_2+\lambda_3) \, \hat n_{E_b}  
  + (\epsilon_b- \lambda_2-\lambda_3) \, \hat n_{F_{2b}} 
\nonumber\\
&& + 2\sqrt{6} \, \lambda_4 \, \left[ 
(s^{\dagger}_{F_2} b_{F_2})^{A_1} + \mbox{h.c.} \right] 
+ 2\sqrt{6} \, \lambda_5 \, 
\left[ (s^{\dagger}_{A_1} b_{A_1})^{A_1} + \mbox{h.c.} \right] ~.
\label{h1}
\ea
There are two terms that contain the spurious $A_1$ bending mode. 
This shows that, in order to remove the spurious interactions from 
$\hat H_1$, the condition $\lambda_5=0$ of Eq.~(\ref{cond1}) is not 
sufficient. In the present symmetry adapted analysis we find two 
conditions: $\lambda_5=0$ and $\epsilon_b+5\lambda_2+\lambda_3=0$~. 
We remark that, since the expectation value of $\hat n_{A_{1b}}$ 
in the physical basis vanishes, the omission of the second 
condition has no further consequences. 

A similar analysis 
can be carried out for the interactions in $\hat H_{2s}$ and 
$\hat H_{2b}$. The operators in Eq.~(\ref{ham2})
represent strong local interactions, and their representation 
in terms of the normal operators of Eqs.~(\ref{stretch}) 
and~(\ref{bend}) acquire a complicated form which involves 
operators that transfer quanta between different modes. 
The Hamiltonian $\hat H_{2s}$ only depends on the stretching 
modes and hence does not contain spurious interactions. 
However, the bending contribution $\hat H_{2b}$ contains 
terms that transfer quanta between spurious and physical states. 
This can be seen by calculating its matrix elements in the 
basis states with $A_1$ symmetry that have two quanta in the 
bending vibrations 
\ba
| \Gamma \rangle &=& \frac{1}{\sqrt{2}} 
(b^{\dagger}_{\Gamma} b^{\dagger}_{\Gamma})^{A_1} | 0 \rangle ~, 
\ea
with $\Gamma=A_1$, $E$ and $F_2$, respectively.  
The states with $\Gamma=E$, $F_2$ belong to the physical basis 
of Eq.~(\ref{basis}), whereas the state with $\Gamma=A_1$ is spurious. 
In this basis the matrix elements of $\hat H_{2b}$ are 
\ba
-X_b \left( \begin{array}{ccc} 
\frac{1}{3} & \frac{\sqrt{2}}{3} & \frac{1}{\sqrt{3}} \\
\frac{\sqrt{2}}{3} & \frac{2}{3} & \sqrt{\frac{2}{3}} \\
\frac{1}{\sqrt{3}} & \sqrt{\frac{2}{3}} & 1 
\end{array} \right) ~. 
\ea
The nonvanishing matrix elements in the first column and the 
first row show that 
$\hat H_{2b}$ contains terms which involve the spurious $A_1$ 
bending mode. For this reason the parameter $X_b$ in Eq.~(\ref{ham2}) 
must vanish in order to remove the spurious contributions from the 
Hamiltonian, in agreement with the result obtained in \cite{Ma}. 
However, this condition also removes the physical part of 
$\hat H_{2b}$ that couples the states with $\Gamma=E$ and 
$F_2$.

In summary, in this section we have shown that in order to remove 
the spurious interactions from the Hamiltonian of \cite{Ma} three 
conditions have to be fulfilled
\ba
\lambda_5 \;=\; 0 ~, \hspace{1cm} X_b \;=\; 0 ~,
\hspace{1cm} \epsilon_b+5\lambda_2+\lambda_3 \;=\; 0 ~.
\label{cond2}
\ea
In the calculation of the vibrational spectrum of methane 
presented in \cite{LF}, no stretching-bending interactions 
were included ($X_b=0$). The spurious components in the Hamiltonian 
were removed by requiring that its expectation value in the 
spurious state with one quantum in the $A_1$ bending mode vanishes. 
This method only leads to exact results in the harmonic limit 
in which it reduces to the last condition of Eq.~(\ref{cond2}). 

An additional comment concerns the basis used in \cite{Ma}. The basis 
functions in general do not carry the quantum numbers that correspond 
to the number of quanta in each of the normal vibrational modes. 
This represents a serious disadvantage in the process of identifying 
the calculated energies with the experimental ones, which traditionally 
are reported in terms of these labels. 
This problem was overcome in \cite{LF} by constucting the physical basis 
by means of the ${\cal T}_d$ Clebsch-Gordan coefficients. We note, 
however, that in Eq.~(3.12) of \cite{LF} the normalization factor 
\ba
| n \rangle |m \rangle &=& \prod_{\mu} \left[ 
\frac{(n_{\mu}+m_{\mu})!}{n_{\mu}!m_{\mu}!} \right]^{1/2} | n+m \rangle ~, 
\label{norm}
\ea
was missing. When this factor is included, the coupling procedure provides 
an orthonormal symmetry-adapted basis for the physical space with the 
desired labels ($\nu_1,\nu_2,\nu_3,\nu_4$). 

\section{Symmetry adapted approach}

At this point we would like to mention that recently the $U(2)$ 
model used in \cite{LF} has been further developed 
\cite{X3,Be4,ozone,metano} by elaborating a 
tensorial formalism analogous to the one suggested above. 
In this symmetry adapted approach the local $U_i(2)$ algebras 
$\{ \hat G_i \} \equiv \{ \hat N_i, \, \hat J_{+,i}, 
\, \hat J_{-,i}, \, \hat J_{0,i} \}$ are projected to 
${\cal T}_d$ tensors 
\ba
\hat T^{\Gamma_s}_{\mu,\gamma} \;=\;
\sum_{i=1}^{4} \, \alpha^{\Gamma,\gamma}_i \, \hat J_{\mu,i} ~,
\hspace{1cm} 
\hat T^{\Gamma_b}_{\mu,\gamma} \;=\;
\sum_{i=5}^{10} \, \beta^{\Gamma,\gamma}_i \, \hat J_{\mu,i} ~,
\label{tensor}
\ea
where $\mu=\pm,0$. The coefficients have the same 
meaning as in Eqs.~(\ref{stretch}) and~(\ref{bend}). The ${\cal T}_d$ 
invariant Hamiltonian is constructed by repeated couplings of these 
tensors to a total symmetry $A_1$. For example, the interactions that 
are at most quadratic in the generators 
and conserve the total number of quanta are given by 
\ba
\hat{\cal H}_{\Gamma_x} &=& \frac{1}{2N_{x}} \sum_{\gamma} \left( 
  \hat T^{\Gamma_x}_{-,\gamma} \, \hat T^{\Gamma_x}_{+,\gamma}
+ \hat T^{\Gamma_x}_{+,\gamma} \, \hat T^{\Gamma_x}_{-,\gamma} 
\right) ~,
\nonumber\\
\hat{\cal V}_{\Gamma_x} &=& \frac{1}{N_{x}} \sum_{\gamma} \,
\hat T^{\Gamma_x}_{0,\gamma} \, \hat T^{\Gamma_x}_{0,\gamma} ~.
\nonumber\\
\hat{\cal H}_{sb} &=& \frac{1}{2\sqrt{N_sN_b}} \sum_{\gamma} \left( 
  \hat T^{F_{2,s}}_{-,\gamma} \, \hat T^{F_{2,b}}_{+,\gamma}
+ \hat T^{F_{2,s}}_{+,\gamma} \, \hat T^{F_{2,b}}_{-,\gamma} \right) ~,
\nonumber\\
\hat{\cal V}_{sb} &=& \frac{1}{\sqrt{N_sN_b}} \sum_{\gamma} \, 
\hat T^{F_{2,s}}_{0,\gamma} \, \hat T^{F_{2,b}}_{0,\gamma} ~. 
\label{hint}
\ea
Here $\Gamma=A_1$, $F_2$ for stretching vibrations ($x=s$) and 
$\Gamma=E$, $F_2$ for bending vibrations ($x=b$). 
The harmonic limit of these interactions is found by taking 
the limit $N_i \rightarrow \infty$, so that \cite{X3,Be4} 
\ba
\lim_{N_i \rightarrow \infty} 
\frac{\hat J_{+,i}}{\sqrt{N_i}} \;=\; a_i ~,
\hspace{1cm} 
\lim_{N_i \rightarrow \infty} 
\frac{\hat J_{-,i}}{\sqrt{N_i}} 
\;=\; a^{\dagger}_i ~,
\hspace{1cm} 
\lim_{N_i \rightarrow \infty} 
\frac{2\hat J_{0,i}}{N_i} \;=\; 1 ~,
\ea
where $a_i$ and $a^{\dagger}_i$ are the boson operators 
used in \cite{Ma}. In this harmonic limit the interactions of 
Eq.~(\ref{hint}) take the form
\ba
\lim_{N_s \rightarrow \infty} \, \hat{\cal H}_{\Gamma_s} 
&=& \frac{1}{2} \sum_{\gamma} \left( 
  s^{\dagger}_{\Gamma,\gamma} \, s_{\Gamma,\gamma}
+ s_{\Gamma,\gamma} \, s^{\dagger}_{\Gamma,\gamma} \right) ~,
\nonumber\\
\lim_{N_b \rightarrow \infty} \, \hat{\cal H}_{\Gamma_b} 
&=& \frac{1}{2} \sum_{\gamma} \left( 
  b^{\dagger}_{\Gamma,\gamma} \, b_{\Gamma,\gamma}
+ b_{\Gamma,\gamma} \, b^{\dagger}_{\Gamma,\gamma} \right) ~,
\nonumber\\
\lim_{N_{x} \rightarrow \infty} \, \hat{\cal V}_{\Gamma_x} &=& 0 ~,
\nonumber\\
\lim_{N_s, N_b \rightarrow \infty} \, \hat{\cal H}_{sb} 
&=& \frac{1}{2} \sum_{\gamma} \left( 
  s^{\dagger}_{F_2,\gamma} \, b_{F_2,\gamma}
+ s_{F_2,\gamma} \, b^{\dagger}_{F_2,\gamma} \right) ~,
\nonumber\\
\lim_{N_s, N_b \rightarrow \infty} \, \hat{\cal V}_{sb} &=& 0 ~.
\label{harlim}
\ea
which corresponds to the interactions appearing in Eq.~(\ref{h1}), 
with the exception of the spurious interaction which, by construction, 
does not arise in this approach \cite{metano}. 
The symmetry adapted basis states are constructed as in \cite{LF}, 
but now taking into account the appropriate normalization coefficients 
of Eq.~(\ref{norm}). The application of this model to the vibrational 
spectrum of methane yields a fit which is an order of magnitude more 
accurate than those of \cite{Ma} and \cite{LF}. 

\section{Summary and conclusions}

Summarizing, we have analyzed the boson realization model presented 
by Zhong-Qi Ma et al. in \cite{Ma} and found the following results: 

(i) The approach presented in \cite{Ma} does not remove all 
spurious interactions from the Hamiltonian. Moreover, the condition 
$X_b=0$ eliminates, in addition to spurious interactions, also some 
physical interaction terms. 

(ii) Although the method proposed to remove the spurious states gives 
the correct results, it becomes increasingly difficult 
to apply when additional higher order interactions have to be included 
in the Hamiltonian. 

(iii) In general, the basis in which the Hamiltonian is diagonalized 
does not carry the quantum numbers associated to the number of quanta 
in each fundamental vibrational mode. This problem can be solved either 
by using the ${\cal T}_d$ Clebsch-Gordan coefficients or by 
diagonalizing the number operators of Eq.~(\ref{number}). 

(iv) The formulation in terms of tensor operators provides a natural
way to eliminate the spurious interactions from the Hamiltonian. 
Besides, we have shown that the approach of \cite{Ma} corresponds 
to the harmonic limit of a more general method based on coupled 
$U(2)$ algebras \cite{X3,Be4,ozone,metano}.

\section*{Acknowledgements}

This work was supported in part by the 
European Community under contract nr. CI1$^{\ast}$-CT94-0072, 
DGAPA-UNAM under project IN105194, CONACyT-M\'exico under project 
400340-5-3401E and Spanish DGCYT under project PB95-0533.

\end{document}